\documentclass{article}

\usepackage[nonatbib, final]{neurips_2025}

\usepackage[utf8]{inputenc} 
\usepackage[T1]{fontenc}    
\usepackage{hyperref}       
\usepackage{url}            
\usepackage{booktabs}       
\usepackage{amsfonts}       
\usepackage{graphicx}      
\usepackage{nicefrac}       
\usepackage{microtype}      
\usepackage{xcolor}         
\usepackage{svg}
\usepackage[babel]{csquotes}
\usepackage{amsmath}
\usepackage{booktabs}
\usepackage{makecell}   
\usepackage{siunitx}    
\usepackage{svg}

\usepackage{xcolor}
\usepackage{mdframed}

\usepackage{csquotes}
\setquotestyle[guillemets]{german}

\usepackage[style=numeric, backend=biber, natbib=true,maxcitenames=2]{biblatex} 
\title{Solver-in-the-Loop Applications in Astrophysical (Magneto)hydrodynamics}

\author{%
  Leonard Storcks \quad \texttt{leonard.storcks@iwr.uni-heidelberg.de}
  \and
  Tobias Buck \quad \texttt{tobias.buck@iwr.uni-heidelberg.de} \\
  AstroAI Lab, Interdisciplinary Center for Scientific Computing \\
  Heidelberg University
}

\begin{document}

\maketitle

\begin{abstract}
We present two promising applications of training machine learning models inside a differentiable astrophysical (magneto)hydrodynamics simulator. First, we address the problem of slow convergence in hydrodynamical simulations of wind-blown bubbles with radiative cooling. We demonstrate that a learned cooling function can recover high-resolution dynamics in low-resolution simulations. Secondly, we train a convolutional neural network to correct 2D magnetohydrodynamics simulations of a specific blast wave problem. These case studies pave the way for the principled application of more general machine learning models inside astrophysical simulators. The code is available open source under \url{https://github.com/leo1200/eurips25corr}.
\end{abstract}

\section{Introduction}

From a single star to galactic or even cosmological scales, astrophysical fluid simulators are required to encompass multiple physical effects on a wide range of spatial and temporal scales \citep{teyssier15}. Limitations in compute and memory constrain the resolutions possible at small scales. Under-resolving small-scale dynamics can impact the overall accuracy of a simulation. For instance, under-resolving the feedback regions around stars can lead to inaccurate momentum input \citep{lancaster24, vanMarle2011}. This necessitates the development of subgrid models \citep{stinson2006}.

Machine learning techniques and tools are opening new avenues for developing subgrid models. For instance \citet{hirashima25} learn a surrogate model for supernovae and replace regions of interest in the simulation with the surrogate results. However, extracting and replacing whole regions in the simulation is not suitable for every application. For example, under-resolved cooling results in inaccurate dynamics throughout the simulation \citep{vanMarle2011}.

Differentiable simulators open another option: machine learning models can be trained inside the simulator to effectively model unresolved dynamics. Such solver-in-the-loop techniques\footnote{Solver-in-the-loop refers to the integration of the numerical solver, i.e. the differentiable simulator, into the training loop of a machine learning model.} \citep{um21} allow for the trained model to be aware of its interaction with the simulator.

Traditional astrophysical solvers like \citep{Springel10,Stone08,Teyssier2002, shamrock25} are not differentiable
and differentiable fluid solvers like \citep{jaxfluids2,xlb} lack the physics required for astrophysical applications. In astrophysical gases, magnetic fields, self-gravity, energy losses through radiation (radiative cooling) etc. can play important roles. An exception is \texttt{jf1uids}\footnote{https://github.com/leo1200/jf1uids/} \citep{storcks24} the first differentiable magnetohydrodynamical (MHD) simulator of its kind. Using \texttt{jf1uids} we present two novel solver-in-the-loop applications for astrophysics, an effective radiative cooling model restoring high-resolution dynamics on a wind-blown bubble and a corrector for an MHD blast wave problem which retains a divergence-free magnetic field.

The governing set of equations relevant for our test cases is (compare Equation Set 2 in \cite{clarke15} for the MHD part and Eq. 2 in \cite{townsend2009} for the cooling)

\begin{equation}
  \begin{aligned}
    \frac{\partial \rho}{\partial t} & =-\nabla \cdot(\rho \vec{v}) \\
    \frac{\partial \vec{v}}{\partial t} & = -(\vec{v} \cdot \nabla) \vec{v} - \frac{1}{\rho} \nabla p +\frac{1}{\rho}(\nabla \times \vec{B}) \times \vec{B} \\
    \frac{\partial p}{\partial t} & =- \vec{v} \cdot \nabla p -\gamma p \nabla \cdot \vec{v} - (\gamma - 1) \frac{\rho^2}{\mu_e \mu_H} \Lambda(T) \\
    \frac{\partial \vec{B}}{\partial t} & = \vec{\nabla} \times (\vec{v} \times \vec{B})
  \end{aligned}
\end{equation}

with density $\rho$, velocity $\vec{v}$, pressure $p$, magnetic field $\vec{B}$, adiabatic index $\gamma$, effective molecular weights $\mu_e,\mu_H$ and a cooling function $\Lambda(T)$. The temperature is given by $T = \frac{p\mu}{\rho k}$ with mean molecular weight $\mu$ and Boltzmann constant $k$. The cooling function $\Lambda(T)$ (at solar metallicity) is taken from \citet{schure2009}. \texttt{jf1uids} uses a constrained transport scheme based on \citet{pang24} for the MHD, differentiation through a simulation with adaptive time stepping is enabled by optimal online checkpointing \citep{stumm10}.\footnote{The optimal online checkpointing was implemented in \url{https://github.com/patrick-kidger/equinox/blob/main/equinox/internal/\_loop/checkpointed.py} by Patrick Kidger.} For radially symmetric simulations like the first test case the geometric formulation of the Euler equations from \citet{euler_spherical} is applied.


\section{Cooling functions trained to recover high-resolution dynamics}

We consider the radially symmetric stellar wind setup of \citet{vanMarle2011} without magnetic fields ($\vec{B} = \vec{0}$). \citet{vanMarle2011} demonstrate that the evolution of wind-blown bubbles with radiative cooling is highly resolution dependent, converging only at resolutions above approximately $10^5$ cells corresponding to $9$ levels of adaptive mesh refinement in their setup.\footnote{This resolution sensitivity relates to the cooling instability - high density regions cool more rapidly ($\frac{dp}{dt}_{cool} \propto \rho^2\Lambda(T)$), get compressed by the surrounding gas and therefore cool even more rapidly.} This problem is illustrated in Fig. \ref{fig:problem_setting} in the appendix. \citet{vanMarle2011} conclude \enquote{Since the number of gridpoints would quickly become prohibitive if this simulation was carried out on a fixed grid, adaptive mesh refinement can be considered
a necessity for simulations of this kind}. But even for a single star in a full three-dimensional simulation with adaptive mesh refinement, effective resolutions comparable to the radial one-dimensional cases will be prohibitive. We propose an alternative cost-efficient approach which works on low-resolution static grids: An effective cooling function $\Lambda_{\Phi}(T)$ parameterized by a neural network (in our case a multilayer perceptron with three hidden layers and $256$ neurons each) is trained to restore high-resolution dynamics. $\Lambda_{\Phi}(T)$ is pre-trained on the \citet{schure2009} cooling.

\begin{figure}[!htb]
  \centering
  \includegraphics[width=0.7\textwidth]{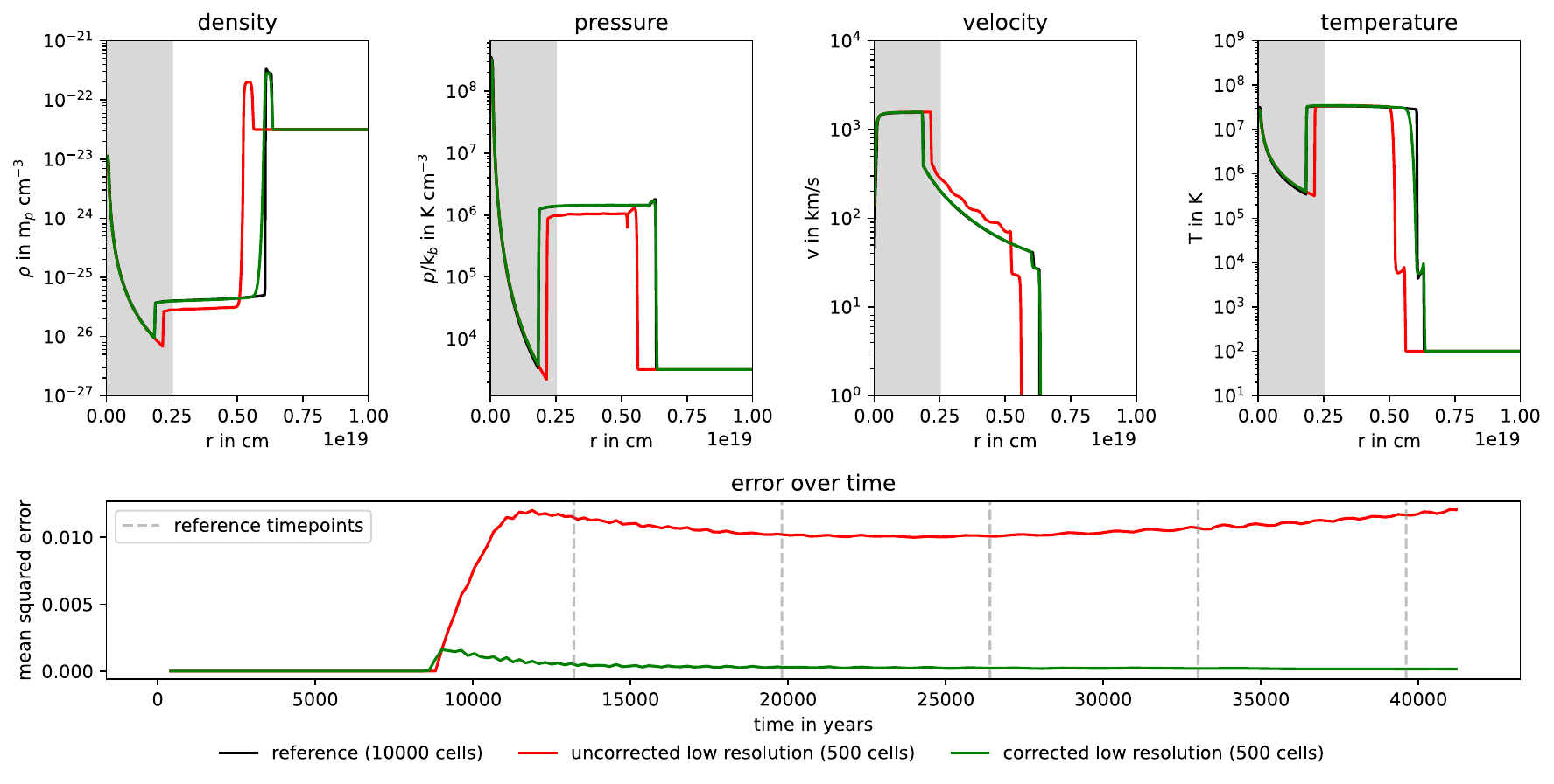}\hfill
  \caption{Radial profiles of density, pressure, velocity, and temperature (top row) comparing the high-resolution reference simulation ($ N = 10000$ cells), the uncorrected low-resolution simulation ($N = 500$ cells), and the corrected low-resolution simulation using the learned effective cooling function. The shaded region is ignored for reasons discussed in Sec. \ref{app:injection}. The bottom panel shows the mean squared error between the low-resolution and high-resolution solutions over time.}
  \label{fig:profile_comparison}
\end{figure}

\begin{figure}[!htb]
  \centering
  \includegraphics[width=0.6\textwidth]{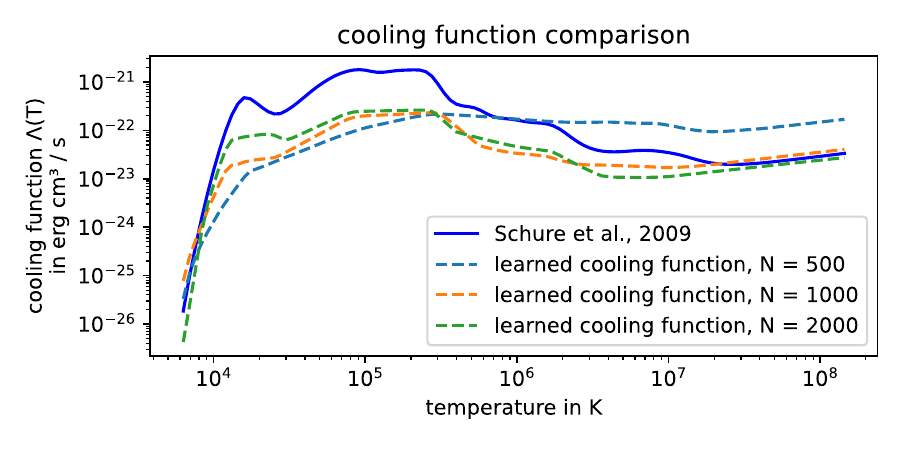}\hfill
  \caption{Comparison between the radiative cooling function from \citet{schure2009} and the learned effective cooling functions for simulations at different resolutions. The learned functions compensate for under-resolved high-resolution details.}
  \label{fig:cooling_function_comparison}
\end{figure}

The optimization goal is given by

\begin{equation}
\begin{gathered}
    \phi^{\star}=\arg \min _\phi \mathcal{L}(\phi)=\arg \min _\phi\left\langle\left\|\frac{\mathbf{U}_{i}(\phi, t)-\mathbf{U}_{i}^{\mathrm{ref}}(t)}{\mathbf{U}_{\max }^{\mathrm{ref}}}\right\|_2^2\right\rangle_{N_{\text{low-res}}>i>i_{\min }, t \in \mathcal{T}} \\
    \mathcal{T}=\left\{\left.t_j=\frac{j}{N_{\text {snap }}+1} t_{\max }^{\operatorname{train}} \right\rvert\, j=2,3, \ldots, N_{\text {snap }}+1\right\}, t_{\max }^{\operatorname{train}} = 1.25\cdot 10^{12} \text{s}, N_{\text {snap }} = 5
\end{gathered}
\end{equation}

where $\mathbf{U}_{i} = (\rho_i, v_i, p_i)$ is the fluid state of cell $i$, the reference $\mathbf{U}_{i}^{\mathrm{ref}}$ is obtained by down-averaging a high-resolution simulation with $N_{\text{high-res}} = 10^4$ and $\mathbf{U}_{\max }^{\mathrm{ref}}$ collects the maximum density, velocity, and pressure across time $t \in \mathcal{T}$ and $i>i_{\min}$ with
$i_{\min } = \lfloor N_{\text{low-res}} / 4\rfloor$. In Sec. \ref{app:injection} we discuss these choices in more detail and additionally consider the case of $N_{\text{high-res}} = 10^5$. For $N_{\text{low-res}} = 500$ taking $150$ optimization steps with ADAM \citep{kingma} takes $\sim 1$ hour on an NVIDIA GeForce RTX 2080 Ti. The resulting radial profiles and the loss over the simulation time are shown in Fig. \ref{fig:profile_comparison}. For $N_{\text{low-res}} = 1000, 2000$ we provide the results in Fig. \ref{fig:profile_comparison1k} and \ref{fig:profile_comparison2k} in the appendix. The corresponding effective cooling curves are shown in Fig. \ref{fig:cooling_function_comparison}. Apart from the higher numerical diffusivity expected at lower resolutions, the effective cooling curves enable the simulator to capture the high-resolution dynamics remarkably well. The existence of such an effective cooling function is already a very interesting result which can benefit future theoretical research on low-resolution cooling behavior and paves the way for more general effective models.

\section{Convolutional neural networks as correctors in magnetohydrodynamics simulations}

Next we consider the two-dimensional MHD blast wave problem of \cite[Sec. 3.3]{arepo_mhd} without cooling ($\Lambda(T) = 0$). We train a corrector improving low-resolution dynamics to mimic high-resolution dynamics for the MHD blast. More specifically, for a low-resolution grid with $64^2$ cells we train a convolutional neural network (CNN) $\mathbf{C}: \mathbb{R}^{7x64x64} \rightarrow \mathbb{R}^{7x64x64}$ ($7$ scalar fluid variables, $\rho,v_x,v_y,B_x,B_y,B_z,p$). To ensure that the CNN does not introduce any magnetic field divergence, we apply a discrete curl transformation to its magnetic field output, and call this divergence-free CNN $\tilde{\mathbf{C}}$. The correction step applied to the state array $\mathbf{U} \in \mathbb{R}^{7\times 64 \times 64}$ after each MHD time-step then reads

\begin{equation}
  \mathbf{\tilde{U}} = \mathbf{U} + \Delta t \, \tilde{\mathbf{C}}(\mathbf{U})
\end{equation}

\begin{figure}[!htb]
  \centering
  \includegraphics[width=0.7\textwidth]{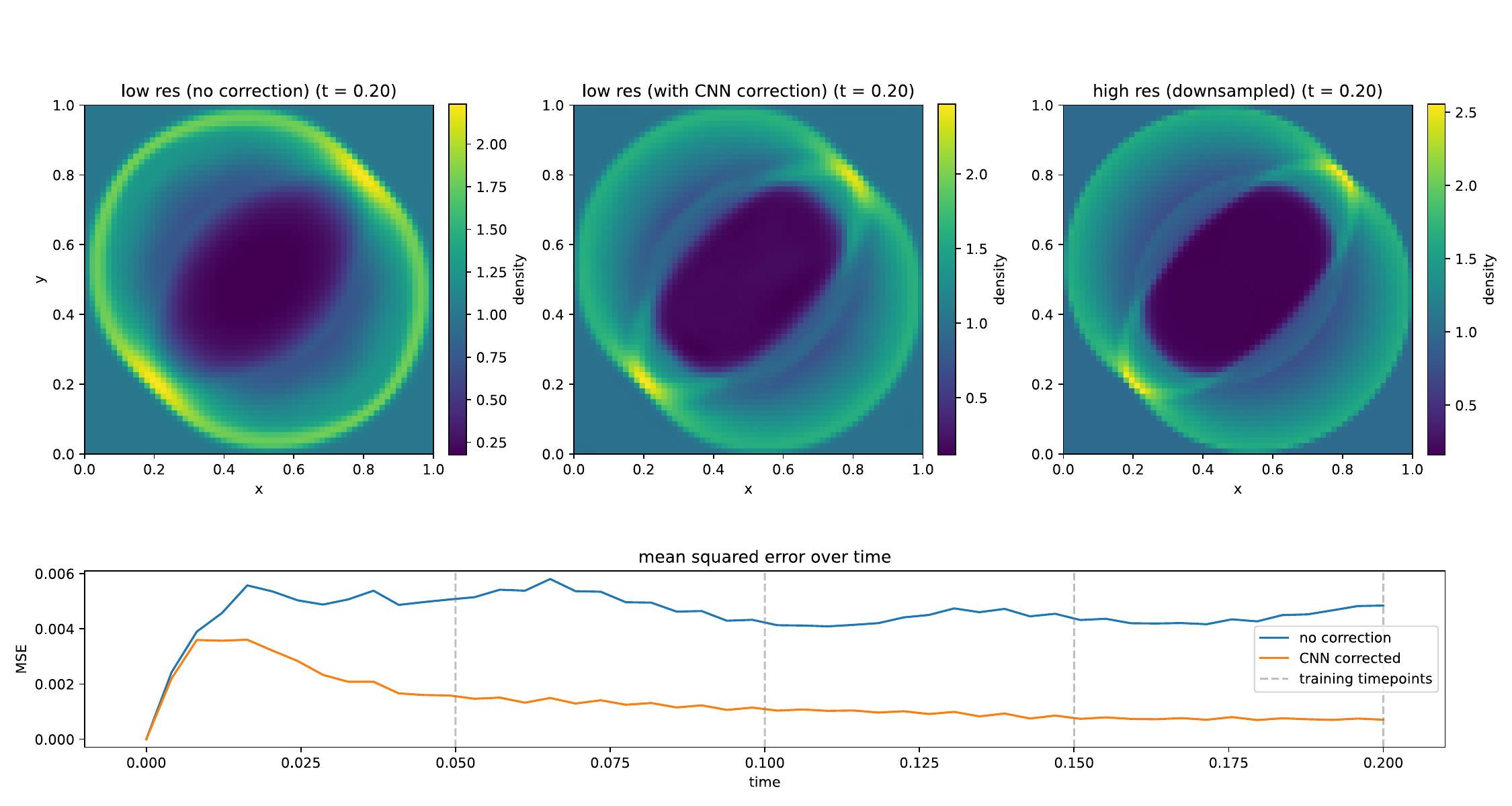}\hfill
  \caption{In the top row from left to right the final density
  field for an MHD blast setup is shown for a low-resolution simulation 
  with $64^2$ cells, a low-resolution with $64^2$ cells but a corrective
  convolutional neural network applied to the state after each time step,
  and a high-resolution simulation down-averaged from $512^2$ to $64^2$ cells.
  In the bottom panel, the mean squared error (MSE) over time of the uncorrected
  and corrected low-resolution simulations with respect to the down-averaged
  high-resolution simulation is shown.}
  \label{fig:blast_correction}
\end{figure}

In our experiment, the CNN consists of a single hidden layer with $16$ channels. We train the CNN inside the simulator against a down-averaged $512^2$ high-resolution reference simulation with $4$ evaluation points in time between 
$t = 0.05$ and $t = 0.2$. The results are shown in Fig. \ref{fig:blast_correction}. In the upper panel, a visual comparison between the uncorrected and corrected low-resolution and reference density fields is shown. The improvement due to the corrector is visually apparent. This visual impression is consistent with the mean squared error (MSE) of the corrected and uncorrected low-resolution results to the reference simulation over time in the lower panel. Until roughly $t = 0.01$ the MSEs are similar, at later times the corrected simulation features errors only a fraction of those from the uncorrected simulation. One epoch of training took $\sim 8.8$ seconds on an NVIDIA GeForce RTX 2080 Ti and we trained for $2000$ epochs. The training loss over epochs is given in Fig. \ref{fig:mhd_loss} in the appendix.

\section{Conclusion and outlook}

We presented two promising solver-in-the-loop applications in astrophysical fluid dynamics, specifically machine learning models trained inside the astrophysical fluid simulator \texttt{jf1uids} to improve low-resolution dynamics. Firstly, we experimentally demonstrated that there exists a learnable effective cooling function which can recover high-resolution dynamics in low-resolution radially symmetric stellar wind simulations with radiative cooling, challenging the need for adaptive mesh refinement as stated by \citet{vanMarle2011} regarding this specific problem. Secondly, we trained an explicit corrector term for two-dimensional MHD blast wave dynamics retaining a divergence-free magnetic field. This corrector only operates on the current fluid state. To our knowledge, our MHD corrector is the first published application of automatic differentiation through a magnetohydrodynamical simulation.

The next step is to generalize such models across resolutions and test problems. We've shown that a specific cooling function can lead to improved dynamics. In the future, a neural operator approach might be able to provide problem-specific cooling functions. Additionally, it would be interesting to transfer the effective cooling function approach to wind-blown bubbles in turbulent media with and without magnetic fields, potentially alleviating the convergence challenges reported by \citet{lancaster24}. We would also like to highlight that machine-learning-based adaptations of physical terms already present in the solver, like our effective cooling function, are easy to integrate into existing non-differentiable simulators after training. Concluding, we hope that these experiments can motivate further research regarding the benefits of differentiable simulators for astrophysical (magneto)hydrodynamics.

\newpage

\printbibliography[title={References}]

\newpage

\appendix

\section{Technical Appendices and Supplementary Material}

\subsection{Problem setting of slow convergence of wind-blown bubble simulations with radiative cooling}

\begin{figure}[!htb]
  \centering
  \includegraphics[width=1.0\textwidth]{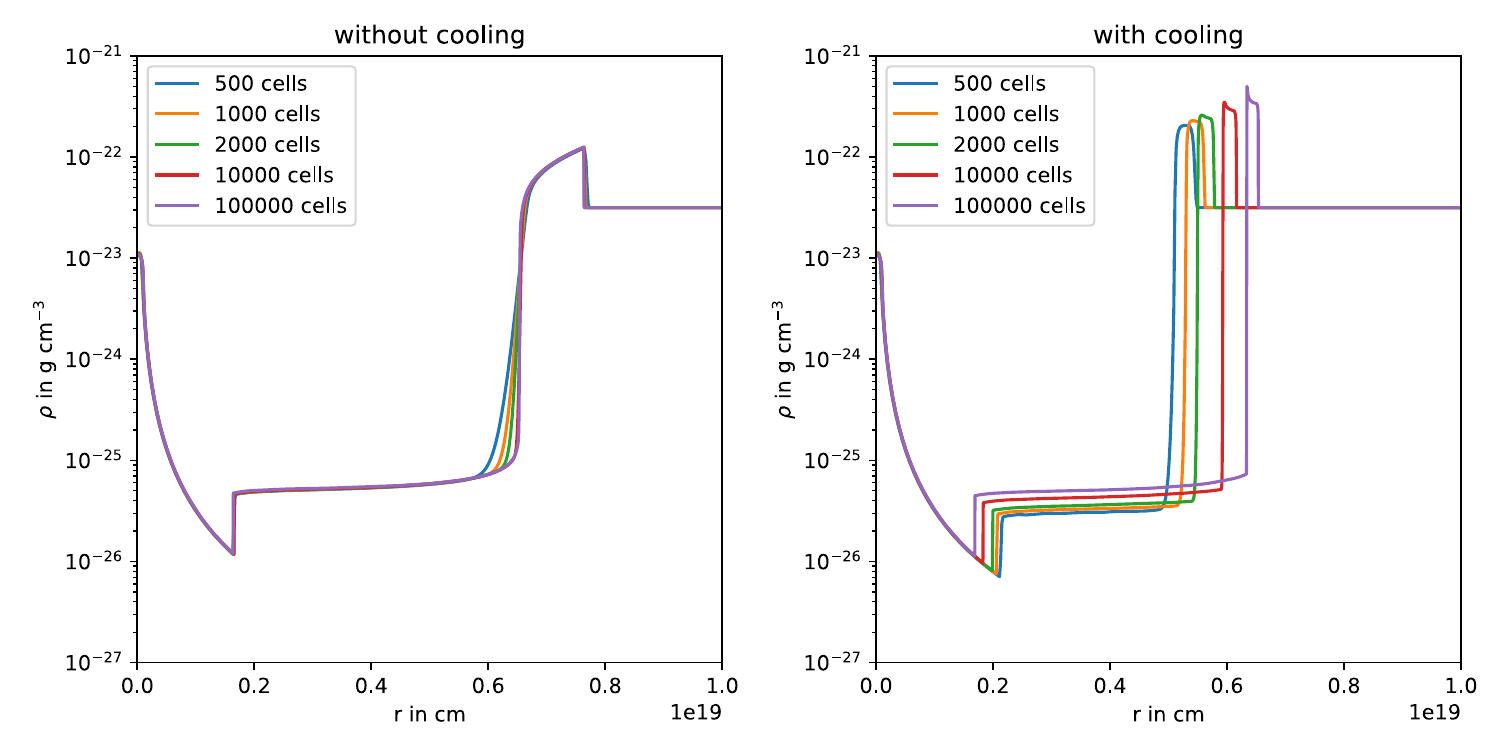}\hfill
  \caption{Density profiles for stellar wind simulations following the setup of \citet{vanMarle2011} at $t = 1.25 \cdot 10^{12} \, \text{s}$ without (left panel) and with cooling (right panel) for different resolutions of $N = \{ 500, 1000, 2000, 10000 \}$ cells. The position of the shock and the profile of the swept-up gas are largely independent of the resolution in the case without cooling but only converge slowly if radiative cooling is introduced \citep{vanMarle2011}.}
  \label{fig:problem_setting}
\end{figure}

Density profiles for wind-blown bubbles with and without radiative cooling at different resolutions are shown in Fig. \ref{fig:problem_setting}.

\subsection{Corrected cooling dynamics at different resolutions}

\begin{figure}[!htb]
  \centering
  \includegraphics[width=1.0\textwidth]{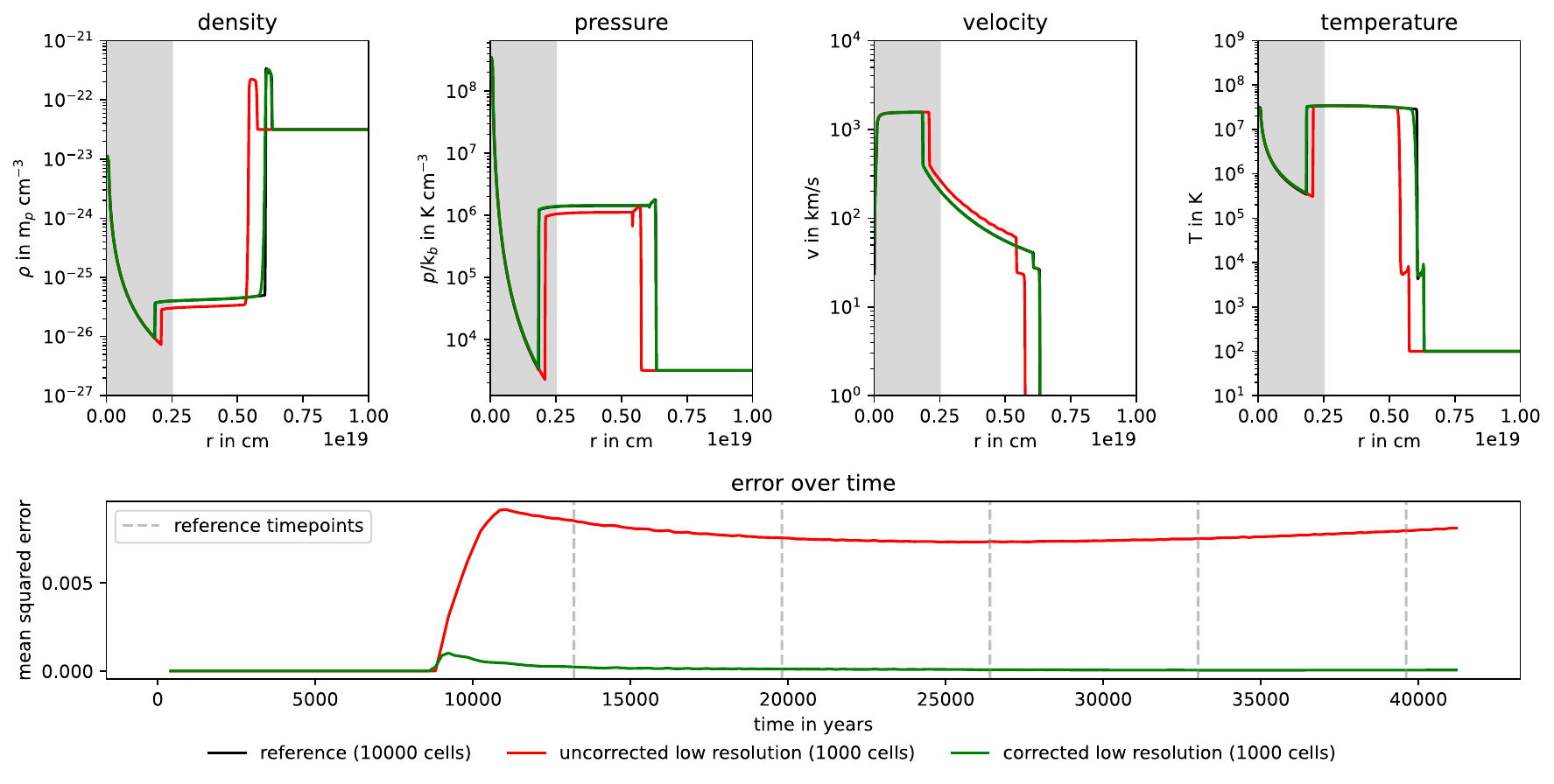}\hfill
  \caption{Radial profiles of density, pressure, velocity, and temperature (top row) comparing the high-resolution reference simulation ($ N = 10000$ cells), the uncorrected low-resolution simulation ($N = 1000$ cells), and the corrected low-resolution simulation using the learned effective cooling function. The bottom panel shows the mean squared error between the low-resolution and high-resolution solutions over time.}
  \label{fig:profile_comparison1k}
\end{figure}

\begin{figure}[!htb]
  \centering
  \includegraphics[width=1.0\textwidth]{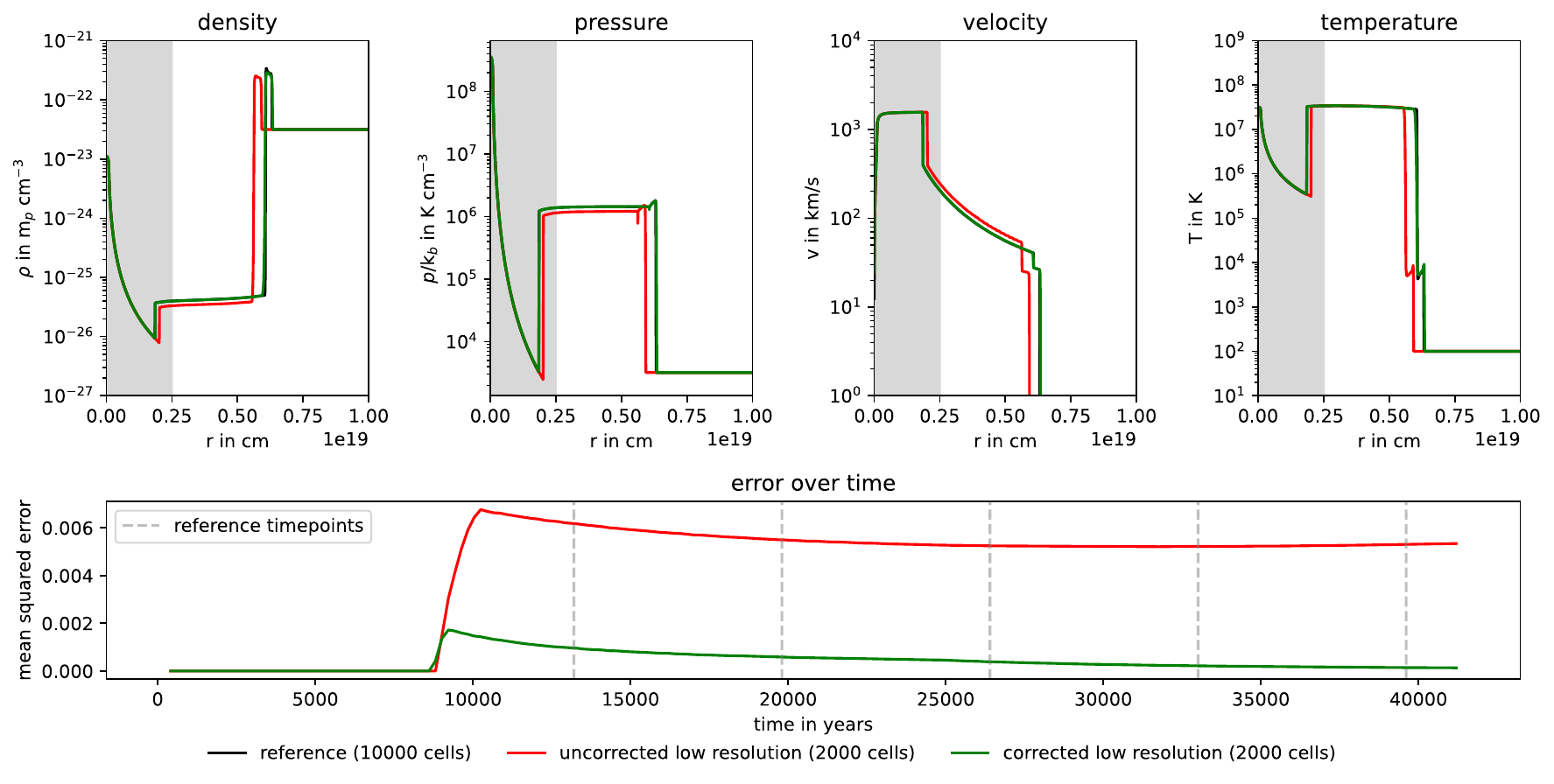}\hfill
  \caption{Radial profiles of density, pressure, velocity, and temperature (top row) comparing the high-resolution reference simulation ($ N = 10000$ cells), the uncorrected low-resolution simulation ($N = 2000$ cells), and the corrected low-resolution simulation using the learned effective cooling function. The bottom panel shows the mean squared error between the low-resolution and high-resolution solutions over time.}
  \label{fig:profile_comparison2k}
\end{figure}

In addition to the corrected results of Fig. \ref{fig:profile_comparison} for $N_{\text{low-res}} = 500$ we show the results for $N_{\text{low-res}} = 1000$ in Fig. \ref{fig:profile_comparison1k} and $N_{\text{low-res}} = 2000$ in \ref{fig:profile_comparison2k}.

\subsection{Experimental setup choices in the cooling correction}
\label{app:injection}

\begin{figure}[!htb]
  \centering
  \includegraphics[width=1.0\textwidth]{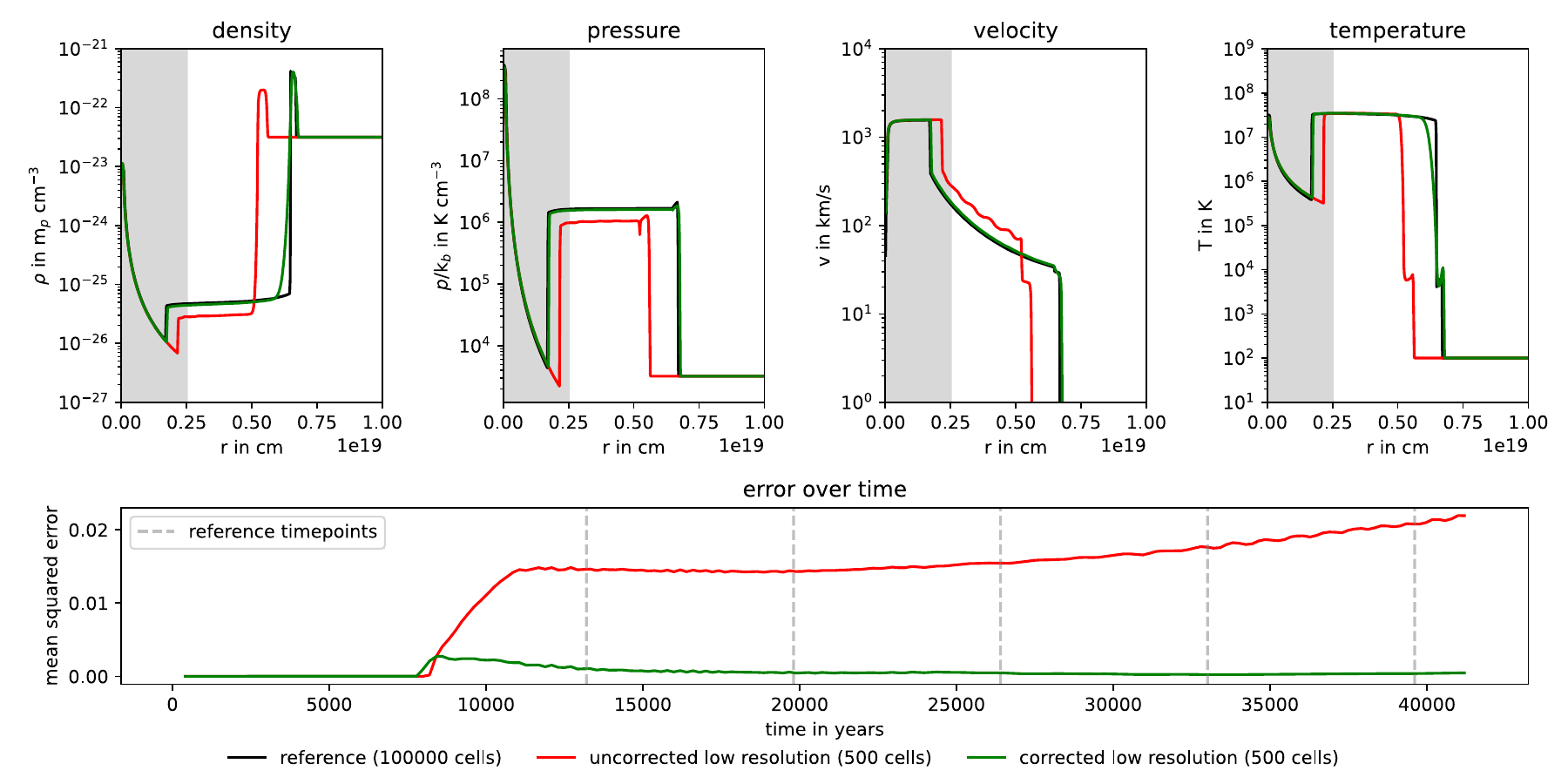}\hfill
  \caption{Radial profiles of density, pressure, velocity, and temperature (top row) comparing the high-resolution reference simulation ($ N = 10000$ cells), the uncorrected low-resolution simulation ($N = 2000$ cells), and the corrected low-resolution simulation here using a temperature and density dependent learned cooling function. Adding the density simplified the training, but we think that by adapting the training procedure or network architecture a purely temperature dependent solution might also be found here.}
  \label{fig:final_comparison_rho}
\end{figure}

We would like to explain two choices here, the resolution of our reference simulations $N_{\text{high-res}} = 10^4$ and the spatial starting index for the loss $i_{\min } = \lfloor N_{\text{low-res}} / 4\rfloor$. A reference resolution of $N_{\text{high-res}} = 10^4$ seemed distinct enough from $N_{\text{low-res}} = 500$ to be interesting while still being cheap (on the order of a minute on an NVIDIA GeForce RTX 2080 Ti). The resolution of convergence $N_{\text{high-res}} = 10^6$ takes on the order of half an hour on the same machine. We are currently looking into $N_{\text{high-res}} = 10^6$ as a reference resolution. Optimization seems to be harder in this case but including the density as a further input to the cooling network might help. This is ongoing research but we provide first results in Fig. \ref{fig:final_comparison_rho}. Regarding $i_{\min } = \lfloor N_{\text{low-res}} / 4\rfloor$ we made this choice as
\begin{itemize}
    \item up to the reverse shock the pressure profile depends on the injection mechanism of the stellar wind, in our case this is always a thermal energy injection into cells within a radius of $0.05 L $ with $L$ being the box length, but it might generally vary
    \item in terms of total volume and mass this innermost region is relatively unimportant (spherical simulation) but weighting with the volume in the loss seemed to put too much weight on the outer regions, so masking this innermost region seemed like a fair compromise
\end{itemize}

\subsection{Training loss of the convolutional neural network corrector for the magnetohydrodynamics simulations}
\begin{figure}[!htb]
  \centering
  \includegraphics[width=1.0\textwidth]{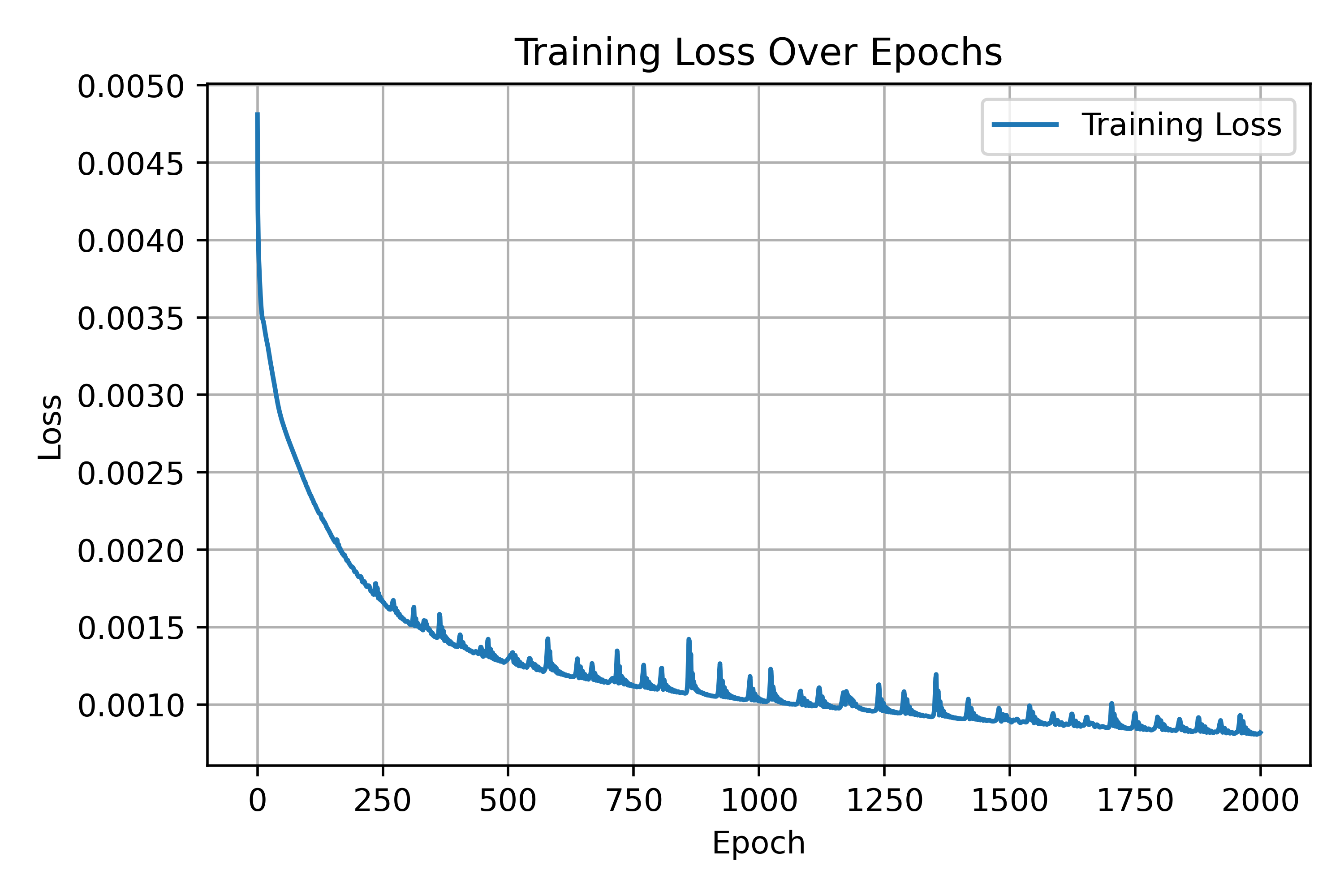}\hfill
  \caption{Training loss over epochs of the convolutional neural network corrector for the magnetohydrodynamic (MHD) blast wave simulation.}
  \label{fig:mhd_loss}
\end{figure}

The training loss over epochs for the MHD corrector is given in Fig. \ref{fig:mhd_loss}.

\end{document}